# High-speed waveguide-coupled graphene-on-graphene optical modulators

Steven J. Koester[1] and Mo Li[2]

*Department of Electrical and Computer Engineering, University of Minnesota-Twin*

*Cities, Minneapolis, MN 55455, USA*

## Abstract

An electro-absorption optical modulator concept based upon a dual-graphene layer is presented. The device consists of a silicon-on-insulator waveguide upon which two graphene layers reside, separated by a thin insulating region. The lower graphene acts as a tunable absorber, while the upper layer functions as a transparent gate electrode. Calculations based upon realistic graphene material properties show that 3-dB bandwidths over 120 GHz (30 GHz) are achievable at near- ($\lambda = 1.55$ μm) and mid- ($\lambda = 3.5$ μm) infrared bands. The effect of background doping and potential fluctuations on the bandwidth, modulation depth and insertion loss are also quantified.

[1] Electronic mail: skoester@umn.edu
[2] Electronic mail: moli@umn.edu

Among graphene's unique electrical, optical, thermal and mechanical characteristics, perhaps its optical properties are the most promising in leading to practical applications in the near future.[1] Many graphene-based optoelectronic devices have been proposed or experimentally demonstrated, including photodetectors,[2] modulators,[3] polarizers[4] and plasmonic structures.[5-7] One of the most intriguing optical properties of graphene is its universal linear absorption coefficient, which equals 2.3% for normal incidence over a broad spectrum range of optical frequency from visible to infrared.[8,9] This absorption is significant considering graphene is only one atomic layer thick. More importantly, interband absorption in graphene at a specific optical wavelength can be tuned by applying a gate voltage to change the carrier concentration and type, due to graphene's gapless band structure and low density of states.[10,11]

This gate-controlled absorption in graphene can be utilized for electro-absorptive modulation for high-speed optical communications. Recently, based on this principle, a graphene modulator integrated on a silicon waveguide operating in the near infrared band was demonstrated and showed modulation frequency up to 1 GHz.[3] In this device, the silicon waveguide itself is utilized as a back-gate to modulate the potential in the graphene. Such a design has the disadvantage that it requires electrical contact to be made through the silicon waveguide and therefore the Si needs to be doped and contacted using a thin, heavily-doped extension region. This increases process complexity and introduces free carriers into the Si waveguide where the optical mode is confined, causing excessive absorption and consequently high insertion loss. In this paper, we propose an alternate waveguide integrated graphene modulator design that is suited for both near-and mid-infrared optical signals. Instead of using the doped silicon waveguide as the gate electrode, we propose to use a second layer of graphene which acts as a transparent gate electrode that modulates absorption in the lower graphene layer. This design approach can significantly improve the modulator speed, reduce the insertion loss and simplify the device design.

The modulator can be fabricated by transferring two layers of CVD-grown graphene[12] on planar single-mode photonic waveguides, as shown in Fig. 1. The first layer of graphene is transferred to form the bottom absorbing layer. To prevent cracking of the graphene at the edge of the waveguide, the substrate can be planarized by depositing an oxide cladding layer using plasma-enhanced chemical vapor deposition (PECVD) or atomic layer deposition (ALD). As shown in the figure, this layer can also provide a spacer between the lower graphene and the top of the waveguide with a thickness of $t_{bot}$. The graphene layer can then be patterned and metal can be deposited to form Ohmic contacts. Similarly, a second layer of graphene which acts as the transparent gate electrode can be transferred after deposition of another oxide layer with thickness of $t_{top}$ and contacts can be formed in a similar fashion to the lower layer. A standard silicon-on-insulator (SOI) substrate can be used to fabricate the waveguide layer for use in a spectral range from $\lambda = 1.1$ μm to mid-infrared with wavelength up to 3.6 μm. At $\lambda > 3.6$ μm the absorption loss in silicon dioxide becomes significant, and therefore, other types of substrates and waveguide structures, such as a suspended silicon membrane, germanium-on-silicon[13] and chalcogenide glasses could be utilized. In the current work, we focus on two specific wavelengths, $\lambda = 1.55$ μm and 3.5 μm using an SOI substrate.

The field amplitude profile in the vertical direction of the waveguide mode is shown in Fig. 1(b). Both the bottom and top graphene layers reside in the evanescent field of the guided optical mode, and the field decay length, $L_z$, is calculated to be 136 nm and 325 nm, for $\lambda = 1.55$ μm and 3.5 μm, respectively. For the modulator to operate properly, however, optical absorption should only be allowed in the bottom layer while the top layer should be transparent at the working wavelength. This situation can be achieved if the background carrier concentration, $n_b$, is within a specified range, and if a DC offset voltage, $V_{DC}$, is applied. Specifically, if the device is tuned such that the bottom-layer Fermi level (relative to the Dirac point) is positioned at $|E_{F-bot}| = \hbar\omega/2$ while in the top-layer, $|E_{F-top}| \gg \hbar\omega/2$, then proper modulator operation is obtained.[10] Here, $\hbar\omega$ is the photon energy. At this condition,[3] the absorption coefficient of graphene is measured to be around 0.09 dB/μm at $\lambda = 1.55$ μm and is expected to be the same for $\lambda = 3.5$ μm.[8]

The modulator performance was simulated using an equivalent circuit model taking into account Fermi-Dirac statistics, quantum capacitance effects, and random potential fluctuations in the graphene. The details of this model are included in the supplemental material,[14] and the main features are summarized below. In this model, the "unit" graphene electron and hole concentrations were determined by integrating the Fermi-Dirac distribution function with the graphene density of states where a random potential, $\delta V_i$, assumed to have a Gaussian distribution with mean value of 0 and standard deviation of $\sigma$, was added to the graphene band structure. A parameter that we call the absorption probability, $P_i$, was defined and can be expressed as

$$P_i = f(-\hbar\omega/2 - E_F + e\delta V_i) \cdot [1 - f(\hbar\omega/2 - E_F + e\delta V_i)], \qquad (1)$$

where $f(E)$ is the Fermi-Dirac distribution function. The average electron and hole sheet densities, $n$ and $p$, as well as the average absorption probability, $P$, were then obtained by averaging over $m = 1000$ random values of $\delta V_i$. The quantum capacitance, $C_Q$, per unit area was calculated from the derivative of the charge concentration with respect to the Fermi-level position.

The procedure above was used to calculate the dependence of $C_Q$, $n$, $p$, $P$ on $E_F$ and the results were tabulated for given values of temperature, $T$, and $\sigma$. For the graphene-on-graphene modulator geometry, the total modulator capacitance, $C_m$, was then calculated as the series combination of the oxide capacitance, $C_{ox}$, and the quantum capacitances of the top and bottom graphene layers, where $C_{ox}$ was calculated using a simple parallel plate approximation with an area, $A = L_m \cdot (W_m + 2W_{over})$ as determined from Fig. 1.

The gate-voltage dependence of the sheet carrier densities in the bottom and top graphene layers was calculated by assuming a fixed background electron concentration, $n_b$, plus the gate-voltage induced charge calculated by integrating $C_m$ over the applied gate voltage. The total series resistance, $R_m$, was calculated by adding the Ohmic resistances of the top and bottom graphene layers as well as the resistance, $R_c$, of the metal contacts to the graphene layers. The mobility, $\mu$, in the graphene layers was assumed to be equal for electrons and holes and was further assumed to be invariant with carrier concentration.[15]

The modulator simulations were performed assuming a gate voltage applied to the top graphene layer, with the bottom graphene grounded. The gate voltage consisted of a DC offset voltage, $V_{DC}$ plus a sinusoidal AC voltage with peak-to-peak amplitude of $V_{osc}$. For all simulations, the DC bias voltage was optimized such that when $V_g = V_{DC}$, then $E_{F-bot} = -\hbar\omega/2$. The 3dB bandwidth, $f_{3dB}$, was calculated from the average value of the RC time constant, $\tau_{RC} = R_m C_m$, over one complete AC voltage cycle.

The absorption in both the top and bottom graphene layers was calculated as the product of the zero-distance absorption coefficient, $\alpha_0$, the vertical decay of the evanescent optical field, and the absorption probability. Therefore, the total absorption coefficient for the bi-layer structure is simply the sum of the absorption in the bottom and top graphene layers:

$$\alpha = \alpha_0 \cdot \left[ P_{bot} e^{-t_{bot}/L_z} + P_{top} e^{-(t_{bot}+t_{top})/L_z} \right]. \tag{2}$$

The relation in (2) corresponds to the total absorption coefficient in the device due to interband absorption, and this value is modulated through the gate voltage-dependence of the absorption probabilities $P_{bot}$ and $P_{top}$. In addition to band-to-band absorption, absorption by the metal contacts was also accounted for by including an additional linear absorption coefficient, $\alpha_m$, so that the total transmission for the length of the modulator can be expressed as $T = e^{-(\alpha+\alpha_m)L_m}$. From this relation, the insertion loss, $L$, and modulation depth, $M$, can be calculated as:

$$L = 10\log(T_{max}), \tag{3}$$

$$M = \frac{T_{max} - T_{min}}{T_{max}}, \tag{4}$$

where $T_{max}$ and $T_{min}$ are the maximum and minimum optical transmission values encountered over the course of one complete AC voltage cycle.

Table I summarizes the design parameters of the modulators for λ = 1.55 μm and 3.5 μm. The distance between the metal contacts and the waveguide was chosen to be a compromise between metal absorption which increases insertion loss and series resistance in the extension regions which reduces bandwidth (Fig. 1c). A conservative overhang width of 0.15 μm for the graphene on the opposite side of the waveguide was chosen, though this value could be reduced further with high resolution lithography. The values chosen for the mobility and contact resistance, while relatively optimistic for typical CVD graphene values, are still conservative relative to the best values reported in the literature for CVD graphene.[16,17] Finally, a peak-to-peak oscillator voltage of 8 V (3.5 V) was chosen for operation at λ = 1.55 μm (3.5 μm).

The simulation results are shown in Fig. 2. The values of $L$, $M$ and $f_{3dB}$ are plotted vs. $n_b$ for λ = 1.55 μm and 3.5 μm. For all simulations, $n_b$ was assumed to be the same for the top and bottom graphene layers. In addition, the results are plotted both for pristine graphene ($\sigma = 0$) and for graphene with a realistic value of potential variations ($\sigma = 65$ mV).[18] The data in Fig. 2 reveals several interesting trends. First, the modulation depth (Fig. 2(a)) depends strongly on the value of $n_b$. At λ = 1.55 μm, values of $M > 0.5$ can be achieved for most background concentrations, except for $-1.76 \times 10^{13}$ cm$^{-2} < n_b < -0.73 \times 10^{13}$ cm$^{-2}$, where $M$ can be as low as 0.1. This behavior is due primarily to absorption in the top graphene layer as will be explained in more detail below. At λ = 3.5 μm, the modulation performance improves, with $M > 0.5$, except over a narrow range of $n_b$. Pristine graphene offers somewhat improved modulation depth compared to disordered graphene. The insertion loss plotted in Fig. 2(b) shows similar behavior to the modulation depth, where at λ = 1.55 µm, significant loss occurs due to absorption in the top graphene layer over a range of $n_b$ values, a window that narrows considerably at λ = 3.5 µm. Nevertheless, acceptable insertion loss values around -2.5 dB are possible at both wavelengths. Though not shown in the figure, improved insertion loss and modulation depth can be achieved by increasing the peak-to-peak gate voltage swing, primarily at the expense of increased power consumption.

Finally, the 3-dB bandwidth, $f_{3dB}$ is plotted in Fig. 2(c). The plot shows that bandwidths as high as 120 GHz are possible for n-type or strong p-type doping. At λ = 1.55 µm, a bandwidth dip for moderate values of $n_b$ occurs when the carrier concentration in the top graphene is at a minimum. The bandwidth at λ = 3.5 µm is roughly 3 times lower than at λ = 1.55 µm primarily due to the larger lateral dimensions of the waveguide and lower carrier concentrations which increase both $R_m$ and $C_m$ compared to shorter wavelengths.

In order to better understand the behavior of the results in Fig. 2, the time-dependent Fermi-level energies in the top ($E_{F-top}$) and bottom ($E_{F-bot}$) graphene layers, for specific values of $n_b$, have been plotted in Fig. 3. In these calculations, the modulator is assumed to be operating at a frequency well below $f_{3dB}$. For each curve, the DC offset voltage has been chosen such that $E_{F-bot} = -\hbar\omega/2$ when $V_g = V_{DC}$. The absorption probability P is depicted in the plot by the shaded background. For both λ = 1.55 µm and λ = 3.5 µm, when the graphene is doped n-type (red lines) or strongly p-type (cyan lines), the Fermi-level in the top graphene remains in a region of low absorption. The peak in the modulation depth (Fig. 2a) occurs when $n_b = 0$, corresponding to the situation where a gate voltage change makes both the top and bottom graphene layers more or less absorbing at the same time (green lines). Conversely, the poorest modulation depth occurs when the starting Fermi-level positions in the top and bottom graphene layers are the same (magenta lines). In this case, the absorption modulation in the bottom layer is counteracted by the opposite effect in the top layer.

A summary of the device performance for a selection of operational parameters is shown in Table 2. At $\lambda = 1.55$ μm, provided very strong p-type doping of $1.8 \times 10^{13}$ cm$^{-2}$ can be realized,[19] a bandwidth of 120 GHz, modulation depth > 0.5 and insertion loss of -2.5 dB can be achieved at $V_{osc} = 8$ V. At $\lambda = 3.5$ μm, the required background hole concentration for optimal operation is reduced to ~ $6.5 \times 10^{12}$ cm$^{-2}$ (which is close to the "natural" p-type doping that occurs in transferred CVD graphene layers).[20] Furthermore, good modulation depth of 0.55 and -2.5 dB insertion loss can be obtained. However, the larger dimensions of the devices at $\lambda = 3.5$ μm restrict the bandwidth to ~ 30 GHz.

The results above provide the basic guidelines for design of graphene-on-graphene electroabsorption optical modulators. However, they do not necessarily predict the ultimate performance capabilities, and improvements in the device design are possible. Since the device operation is limited by RC delays, the utilization of double-sided contacts on the top and bottom graphene layers could improve the bandwidth by reducing the series resistance by nearly a factor of two. If the top graphene background concentration could be controlled independent of the bottom layer, further reduction of the series resistance could be achieved, while at the same time reducing the required value of the DC offset. The use of multi-layer graphene for the top contact could further reduce parasitic series resistance. Improvements in the simulations, including using a distributed RC model, accounting for the effect of disorder on mobility, and consideration of the effect of the graphene on the optical mode[21] could lead to more accurate predictions of the device performance.

In conclusion, an SOI-waveguide-coupled double-layer graphene modulator concept has been described. Bandwidths over 120 GHz (30 GHz) are predicted for a single-ended contact design at $\lambda$ = 1.55 μm (3.5 μm). The effect of background doping and potential fluctuations on the bandwidth, modulation depth and insertion loss have been quantified and suggestions for future design improvements have been provided.

TABLE I. NOMINAL GRAPHENE MODULATOR VALUES USED FOR SIMULATIONS.

| Symbol | Parameter | Default Value |
|---|---|---|
| $\lambda$ | Waveglength | 1.55 μm (3.5 μm) |
| $L_m$ | Modulator length | 60 μm |
| $W_m$ | Waveguide width | 0.5 μm (1.0 μm) |
| $W_{ext}$ | Extension width | 0.4 μm (0.8 μm) |
| $W_{over}$ | Overlap width | 0.15 μm |
| $t_{top}$ | Top oxide thickness | 20 nm |
| $\varepsilon_{top}$ | Top oxide dielectric constant | 3.9 |
| $t_{bot}$ | Bottom oxide thickness | 10 nm |
| $\varepsilon_{bot}$ | Botom oxide dielectric constant | 3.9 |
| $\sigma$ | Random potential fluctuations | 0 and 65 mV |
| $\mu$ | Mobility (both *e* and *h*) | 4000 cm$^2$/Vs |
| $R_c$ | Contact resistance | 400 Ω-μm |
| $T$ | Temperature | 300 K |
| $\alpha_0$ | Zero-distance absorption coefficient | 0.022 μm$^{-1}$ |
| $V_{osc}$ | Peak-to-peak AC voltage | 8 V (3.5 V) |

TABLE II. PERFORMANCE SUMMARY OF GRAPHENE MODULATORS.

| Symbol | Value at $\lambda = 1.55\ \mu m$ | Value at $\lambda = 3.5\ \mu m$ |
|---|---|---|
| $V_{osc}$ | 8 V | 3.5 V |
| $n_b$ | $-1.8 \times 10^{13}\ cm^{-2}$ | $-6.5 \times 10^{12}\ cm^{-2}$ |
| $V_{DC}$ | +5.5 V | +3.7 V |
| $f_{3dB}$ | 121 GHz | 31.8 GHz |
| $M$ | 0.51 | 0.55 |
| $L$ | -2.5 dB | -2.5 dB |

# Figure Captions

Figure 1. (a) Structure of the proposed dual-layer graphene modulator integrated on a waveguide. Overlay shows mode profile of the waveguide's fundamental TE mode. (b) Vertical direction profile of the electric field component amplitude decaying from the top surface of the waveguide. Green shaded area marks the waveguie. Dashed lines mark the graphene layers. (c) Lateral direction profile of the electric field component amplitude decaying from the side surface of the waveguide. Yellow shaded area marks the metal contact.

Figure 2. Plot of (a) modulation depth, (b) insertion loss and (c) bandwidth vs. background electron conentration, $n_b$, for waveguide-coupled graphene-on-graphene optical modulator using the device parameters listed in Table 1. The graphs are shown for both $\lambda = 1.55$ μm (blue) and $\lambda = 3.5$ μm (red), and for pristine graphene with $\sigma = 0$ (dashed lines) as well as graphene with random potential fluctuations of $\sigma = 65$ mV (solid lines). The circles represent the optimal design points listed in Table II.

Figure 3. Plot of Fermi level vs. time for top (colors) and bottom (black) graphene layers for several fixed values of $n_b$ listed on the right side of the graphs. In (a), $\lambda = 1.55$ μm, while in (b), $\lambda = 3.5$ μm. For both (a) and (b), graphene layers with realistic potential fluctuations of $\sigma = 65$ mV have been utilized.

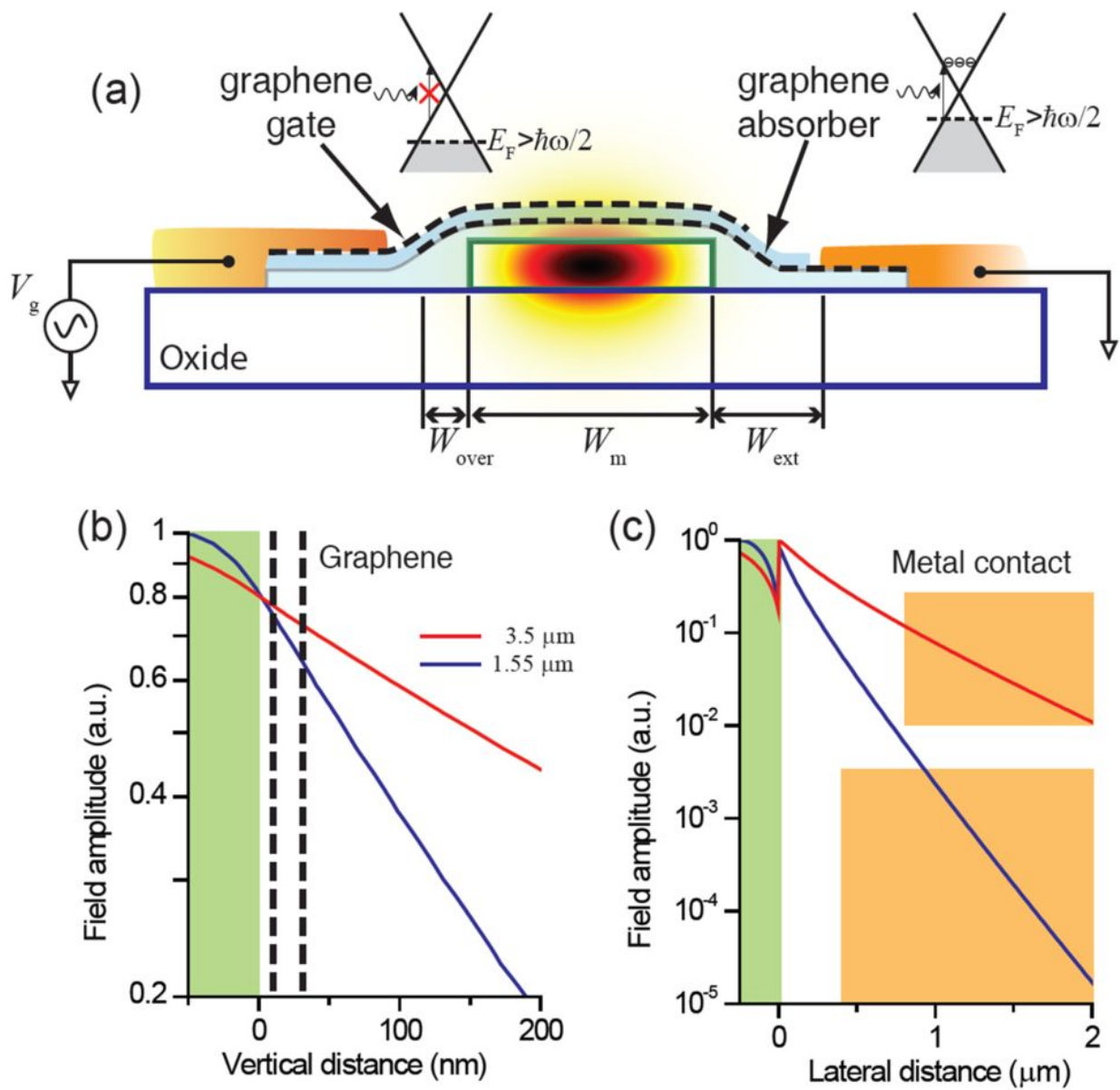

Figure 1, S. J. Koester and M. Li, *Appl. Phys. Lett.*

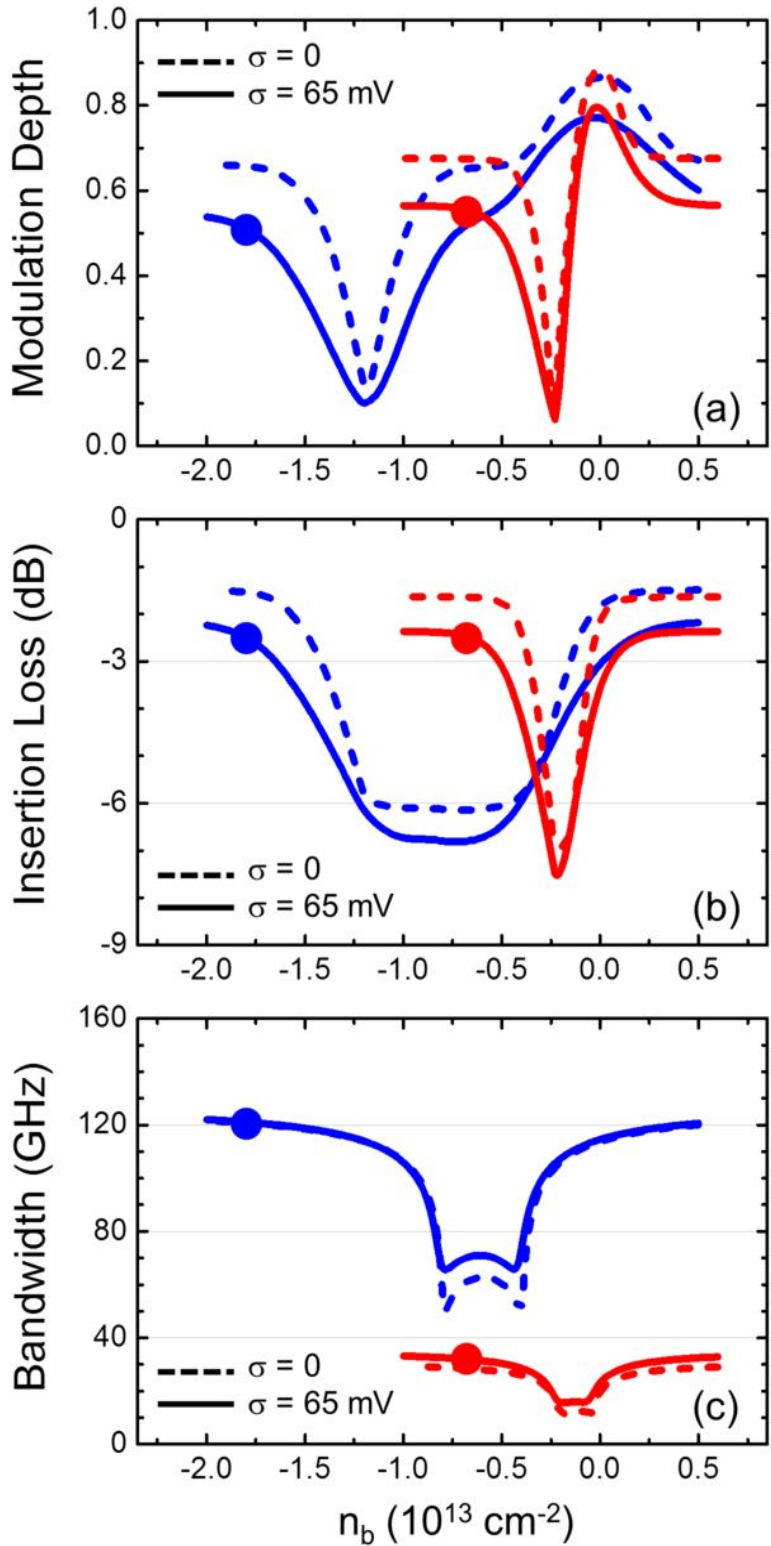

Figure 2, S. J. Koester and M. Li, *Appl. Phys. Lett.*

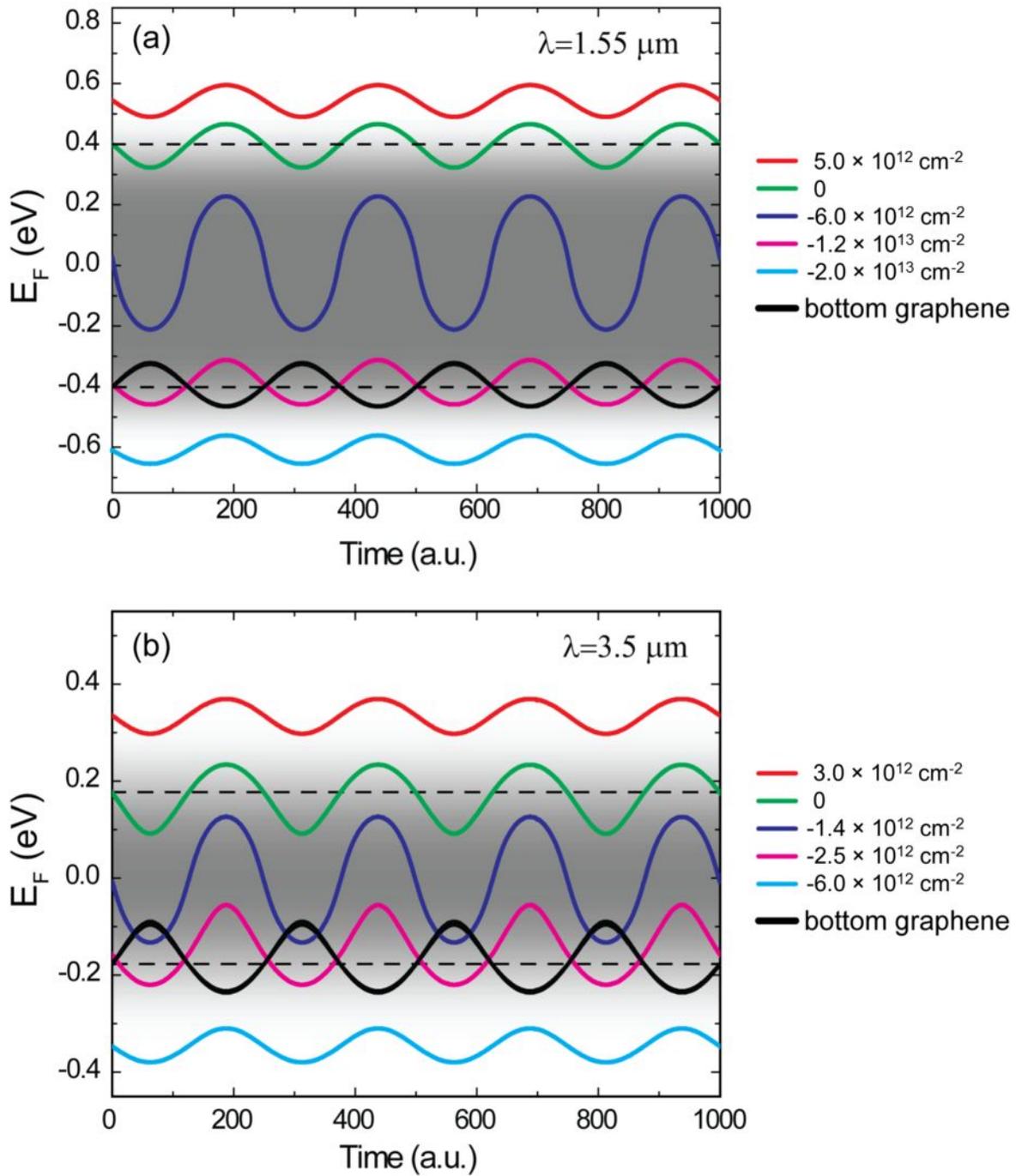

Figure 3, S. J. Koester and M. Li, *Appl. Phys. Lett.*